\documentclass[12pt]{amsart}

\usepackage[lite]{amsrefs}

\usepackage{amssymb}

\usepackage[all,cmtip]{xy}

\usepackage[ruled,vlined,linesnumbered]{algorithm2e}
\let\oldnl\nl
\newcommand{\nonl}{\renewcommand{\nl}{\let\nl\oldnl}}
\usepackage[noend]{algpseudocode}
\usepackage{graphicx}
\usepackage[framemethod=TikZ]{mdframed}
\usepackage{lipsum}
\usepackage{hyperref}
\usepackage{url}
\usepackage{amsmath}   
\usepackage[english]{babel}
\usepackage{subfig}

\usepackage{geometry}

\usepackage{latexsym}
\usepackage{setspace}
\usepackage{cite}
\usepackage{fullpage}
\usepackage[tableposition=above]{caption}
\RequirePackage[english=usenglishmax]{hyphsubst}
\numberwithin{equation}{section}

\theoremstyle{plain} 

\theoremstyle{definition}

\theoremstyle{remark}

\begin{document}

\begin{center}

\title{Data Sharing and Resampled LASSO: A word based sentiment Analysis for IMDb data}

\end{center}

\author{Ashutosh K. Maurya$^{[1,2]}$\\Applied Statistics Unit,\\Indian Statistical Institute, Kolkata}


 \curraddr{}



\date{\today}
\footnotetext[1]{ \* Author: Postal Address: Applied Statistics Unit, Indian Statistical Institute, Kolkata, India, Tel.: +91 9681342717, Email: ashutoshmaurya2010@gmail.com$^{[1]}$}
\footnotetext[2]{\* Project partially supported by Statistical Trainee program, Applied Statistics Unit, ISI, Kolkata and the R. C. Bose Centre for Cryptology and Security, ISI, Kolkata.}
 \begin{abstract}
 In this article we study variable selection problem using LASSO with new improvisations. LASSO uses $\ell_{1}$ penalty, it shrinks most of the coefficients to zero when number of explanatory variables $(p)$ are much larger the number of observations $(N)$. Novelty of the approach developed in this article blends basic ideas behind resampling and LASSO together which provides a significant variable reduction and improved prediction accuracy in terms of mean squared error in the test sample. Different weighting schemes have been explored using \textit{Bootstrapped LASSO}, the basic methodology developed in here. Weighting schemes determine to what extent of data blending in case of grouped data. Data sharing (DSL) technique developed by \cite{gross} lies at the root of the present methodology. We apply the technique to analyze the IMDb dataset as discussed in \cite{gross} and compare our result with \cite{gross}.\\ \\
 
 $\bf Key~ Words:$ High dimensional regression, Sentiment analysis, Weighted LASSO, Resampling, Cross validation, Data sharing.
 \end{abstract}

\maketitle



\section{Introduction}

Extracting important features from any big data is a challenging task. There are number of algorithms for variable selection. In \cite{chen} a notion of data enriched regression has been developed which along with other variable selection techniques (e.g. Ridge\cite{hoerl},  Subset selection\cite{sara}, Least Angle Regression(LARS)\cite{lars}, Least Absolute Shrinkage and Selection Operator(LASSO)\cite{tibshirani} , De-noising\cite{donoho}) which give rise to a host of statistical algorithms for feature selection in big data. For the purpose of this article we study different exploratory variations of regression LASSO based feature selection as developed by \cite{gross}. We introduce two major variations to the DSL algorithm. The first one uses resampling (bootstrap) which provides us with a measure of variation when we use the algorithm by resampling the training sample. The second variation is about choice of weighting scheme across different groups in regularization.


The idea of resampling has a rich history (\cite{SirPC}, \cite{quen1},\cite{quen2}, \cite{tukey},\cite{efron}). Traditionally it comes in two forms: Jackknife and Bootstrap. "The idea of jackknife resampling has evolved into cross validation technique which is commonly used in lasso regularization." The idea of bootstrap, its variations and applications in different context can be found in (\cite{efron},\cite{efron1},\cite{hall}).

The IMDb problem can be viewed as an application of sentiment analysis where the goal is to extract a smaller set of words from very large \textit{bag-of-words} that have better classification of movie ratings. Following\cite{gross} we consider only linear classifiers so that it becomes a linear regression with sparse data. In this study rating have been converted into binary (positive and negative sentiments). This template example can serve as a prototype of a large number of similar problems of sentiment analysis in different areas. The IMDb data used in our analysis consist of 50k movie reviews on a scale of allowing no more than 30 reviews per movie that have been divided into training and test sets of equal size. In our data analysis, we follow the similar resampling procedure to create training and test data sets as outlined in \cite{gross}. It is noteworthy that one can use resampling procedure on the entire datasets, however this will lead to higher computational complexity.

Related literature includes works from Opinion mining, Natural Language Processing, Machine learning among others. Probabilistic classifiers for tracking point of view was designed by \cite{wiebe}.  A machine learning approach to predict the semantic orientation of adjectives (positive or negative) was developed by \cite{hatzivassiloglou}. Substantial research has been done on sentiment analysis. It has been done over various topics such as sentiment analysis over movie reviews \cite{pang}, product reviews (\cite{dave},\cite{na}), news and blogs (\cite{godbole}, \cite{bautin}). Following \cite{pang} conducted an extensive experiment on movie reviews using three traditional supervised machine learning methods (i.e., Naive Bayes (NB), maximum entropy classification (ME), and support vector machines (SVM)). Following \cite{pang}, results indicate that standard machine learning techniques definitively outperform human-produced baselines. However, \cite{pang} found that machine learning methods could not perform as well on sentiment classification as on traditional topic-based categorization.

The article is presented as follows: In Section $(2)$ we present basic theoretical preliminaries about variable selection of high dimensional data. In this section we shall discuss LASSO and other related regularization methods used for this purpose and how we fit our proposed resampling scheme in this framework. In section $(3)$ we illustrate our proposed approaches and methodology for the analysis of the data coming from more than two different sources. We then in section $(4)$ talk about the preparation of the dataset for analysis through an example to apply our methodology. Related experimental results are discussed in section $(5)$ and finally some discussion is given in Section $(6)$.

\section{Preliminaries}

The basic setup starts with a multiple linear regression model for a set of ratings $y_{1},y_{2},...,y_{n}$ with associated explanatory variables $x_{1},x_{2},...,x_{n}$. The explanatory variables are high dimensional that is, $x_{i} = (x_{i1},x_{i2},...,x_{ip})$ where $p>>n$. Moreover we focus on the situation where the design matrix $X= \left(\,(x_{ij})\,\right) $ is binary and sparse.

We consider the multiple linear regression model of $y_{i}$ on $x_{i}$ with intercept. Also while the rating scale is integral valued and bounded which treat $y_{i}$'s as real numbers.

\begin{equation*}
    y_{i} = \mu +  x_{i}^{T}\beta + \epsilon_{i},
\end{equation*}

 $\epsilon_{i}$'s are independently distributed with mean zero and variance $\sigma^{2}$ for $1\leq i \leq n$. and parameter vector $\beta = (\beta_{1},\beta_{2},...,\beta_{p})$.
 
 Furthermore if the data is stratified into $G\geq 2$ different groups we assign a  group indicator variable $g_{i}$ for each observation and thus the regression sum of squares (SS) can be written in the following way
 \begin{equation*}
     \sum_{i=1}^{n}(y_{i}-x_{i}^{T}(\beta + \Delta_{g_{i}}))^{2}.
 \end{equation*}

Following \cite{gross} the coefficients $\Delta= (\Delta_{1},\Delta_{2},\ldots,\Delta_{g})$ represents group effects.

It is well known that ordinary least square (OLS) estimation procedure performs poorly in both prediction and interpretation when ($p>>n$). The unknown intercept and parameter vector can be estimated reasonably well, if $\beta$ is sparse in some sense. The sparsity can be quantified in terms of the $\ell_{q}$ norm for $1\leq q \leq \infty$. The $\ell_{0}$-analogue $\|\beta\|_{0}^{0}$ $=$ $|\{i; \beta_{i}\neq 0 \}|$ (which is not a norm) which counts the number of non-zero entries of the parameter. The notation $\|\beta\|_{0}^{0}$ (where $0^{0}$ = 0)is in analogy to $\|\beta\|_{q}^{q}$ = $\sum_{i=1}^{p}{|\beta_{i}|^{q}}$ for $0<q<\infty$.

To overcome the limitation of OLS in high dimensional setting, we include a regularization term that penalizes the model over coefficients. Subset selection, ridge regression, LASSO and elastic net are important techniques for improving OLS estimates. Subset regression\cite{hall} provides interpretable model. Since it is a discrete process so either variables are retained or dropped form the model hence the results can be extremely variable. Commonly used techniques are based on least squares and a penalty which involves the number of parameters in the candidate sub-model;
\begin{equation*}
    \hat{\beta}(\lambda) = \arg\min_{\beta}{\sum_{i}(y_{i}-x_{i}^{T}\beta)^{2} + \lambda \|\beta\|_{0}^{0} },
\end{equation*}
where the $\ell_{0}$-penalty is $\|\beta\|_{0}^{0}$ $=$ $\sum_{i=1}^{p}1(\beta_{i}\neq 0 )$ . The regularization parameter $ \lambda $ is usually chosen via some cross-validation scheme. 
Under $\ell_{0}$-penalty the estimator is infeasible to compute when p is of large size since the $\ell_{0}$-penalty is a non-convex function of $\beta$\cite{sara}. Many other well known model selection criteria such as the Akaike Information Criterion (AIC), Bayesian Information Criterion (BIC) fall into this framework.

In another approach, \cite{hoerl} have proposed a regression model which minimizes the residual sum of squares subject to a bound on the $L_{2}$- norm of the coefficient. As ridge regression is continuous shrinkage method, it achieves better predictive performance for the case $(p>>n)$. But there is a drawback with ridge regression, it does not set any coefficient to zero and hence does not give any interpretable model\cite{tibshirani}. To overcome with this drawback \cite{tibshirani} proposed a new technique for variable selection when ($p>>n$) called LASSO. The LASSO uses $l_{1}$ penalty which is convex surrogate of l0-norm, results in convex optimization problem and a number of fast algorithms can be used to solve this problem. LASSO shrinks small coefficients to zero and thus performs model selection. In doing so, it results in some bias in the estimated coefficients yet often improves the predictive performance of the model.

The LASSO does poorly for the case $(p>>n)$ or if there are high correlation between predictors\cite{zou}. To deal with this problem \cite{zou} suggested (elastic net) a new regularization and variable selection method which is convex linear combination of $\ell_{1}$ and $\ell_{2}$ norm. The elastic net enjoys property of both ridge and LASSO regression. This regression technique is much useful in the case when ($p>>n$).
Elastic net solves the following problem
\begin{equation*}
  \hat{\beta}_{en} = \arg\min_{\beta}{\sum_{i}(y_{i}-x_{i}^{T}\beta)}^{2} + \lambda P_{\alpha}(\beta),
\end{equation*}

 where
\begin{equation*}
      P_{\alpha}(\beta) = (1-\alpha)\|\beta\|_{2}^{2} + \alpha \|\beta\|_{1},
\end{equation*}
 
is the elastic net penalty\cite{zou}. Which is equivalent to LASSO when ($\alpha$ = 1) and equivalent to ridge regression when ($\alpha$ = 0) and for all $\alpha$ i.e.$ \in [0, 1]$.

Following \cite{chen} proposed a new regularization technique for prediction when the data set come from two different sources. In \cite{chen}, the authors proposes combining two linear regressions on two different datasets, one dataset is used for estimating the regression coefficients, and the squared error penalty term for second dataset is used to improve the estimates of regression coefficients. 
The data enriched regression solves the following problem
\begin{equation*}
      (\hat{\beta},\hat{\Delta})= \arg\min_{(\beta,\Delta)}\sum_{i:g_{i}=1}(y_{i}-x_{i}^{T}\beta)^{2}
      +\sum_{i:g_{i}\neq1}(y_{i}-x_{i}^{T}(\beta+\Delta))^{2} + \lambda P(\Delta),
\end{equation*}
where $\Delta$ is same size as $\beta$ and
\begin{equation*}
     P(\Delta) =\sum_{i:g_{i}=1}(x_{i}^{T}\Delta)^{2},
\end{equation*}
is the  quadratic penalty. Here regularization is done over $\Delta$ only. The above model is not useful to study over different groups together.

In \cite{gross} proposed a new model which is generalization of data enriched regression. The DSL can be used for multiple groups, which are coming from different non-overlapping pre-specified groups. In DSL, the $\ell_{1}$-penalty over both $\beta$ and $\Delta_{g}$ encourages the sparsity and improved estimation in  $\beta$ simultaneously. Suppose we have $n$ observation coming from $G$ groups, and let $p$ be the number of predictors in each group then the generalized data enriched model can be written as
\begin{equation}
    \label{DSLopt}
     y_{i} = x_{i}^{T}(\beta + \Delta_{g_{i}}) + \epsilon_{i},
\end{equation}

where $\epsilon$ is the error term term independent of x and $\beta$.

The DSL minimizes the following
\begin{equation}
    \label{DSL}
     (\hat{\beta},\hat{\Delta_{1}},...,\hat{\Delta_{g}}) = \arg\min \frac{1}{2} \sum_{i} \left(y_{i}-x_{i}^{T}(\beta + \Delta_{g_{i}})\right)^{2} + \lambda \left(\|\beta\|_{1} + \sum_{g=1}^{G}r_{g}\|\Delta_{g}\|_{1}\right)  ,
\end{equation}

where $\lambda$ is the usual regularization parameter. Here $r_{g}$ is used as regularization parameter over groups which controls the amount of sharing between the groups.

\subsection{{\normalfont De-noising} }
To make more precise the interpretation of the extracted features another idea (by removing the noise through regularization) is proposed in this paper. The idea of De-noising was proposed by \cite{donoho}. Following \cite{donoho} suggested thresholding procedure for recovering functions for noisy data. They applied soft thresholding nonlinearity $\eta_{y}(t) = sgn(y)(|y|-t)_{+}$ coordinate-wise to the empirical wavelet coefficient. In \cite{donoho} the chosen threshold is
\begin{equation}
    \label{donoho}
     t_{n} = \sqrt{2log(n)}.\gamma_{1}.\sigma/\sqrt{n}  ,
\end{equation}
where $\gamma_{1}$ is a constant which is defined in \cite{donoho}.

\subsection{{\normalfont Differential Regularization}}
Choice of regularization parameter is a challenging problem in high dimensional data analysis. Regularization has clear benefit in producing sparse solution as well reduces false discovery rate. A larger value of $r_{g}$ accounts for large amount of sharing between the groups and a smaller value of $r_{g}$ accounts there will be less amount of sharing between the groups. Following \cite{gross} suggested some values of $r_{g}$ that is if $\sum_{g}{r_{g}} < 1$ it will be equivalent to separate regression, If  $\sum_{g}{r_{g}} = 1$ then there will be identifiability concern, and a default value of $r_{g}$ which is $1/\sqrt{G}$ when there are same number of observation in each group. From the above conditions it is clear that $\sum_{g}{r_{g}} $ has to be more than one for data shared lasso. In this article we are proposing an idea for choosing the value of $r_{g}$ which results gain in mean squared error than\cite{gross}.


\section{Methodology}

We propose two type of resampling techniques to reduce the size of the dataset ensuring that it conveys similar information concisely. Proposed technique takes care of multi-collinearity, removes redundant features, fastens the time required for performing same computations and it is helpful in noise removal also and as result of that we can improve the performance of models. In addition, we propose new methods of selecting regularization parameters. The data analysis shows the empirical evidence of significance of proposed resampling technique and new method of selecting regularization parameter. We keep the methodology adaptive by taking care of unequal sample sizes in different groups. This will be discussed in greater detail when we analyze the experimental results.

\subsection{Resampling Technique}

In this article we propose two types of resampling schemes,  Bootstrapped LASSO Scheme (BLS) and Bootstrapped Shared LASSO Scheme (BSLS) or Bootstrapped DSL. Here we consider the estimation accuracy of the parameter $\beta$, a different task than prediction. Under compatibility assumptions on the design matrix $X$ and on the sparsity $s_{g_{i},0}$=$|S_{g_{i},0}|$ in a linear model in G groups, let $S_{g_{i},0}$ denote the number of active set of variables initially, defined as
\begin{equation}
    S_{g_{i},0}(\lambda_{g_{i}}) = \{j;\, \beta_{g_{i},j}^{0}\neq 0. \,\,\, j=1,\ldots,p\},
\end{equation}
 and the set of estimated variables using LASSO be given as
\begin{equation}
    \hat{S}_{g_{i}}(\lambda_{g_{i}}) = \{j;\, \hat{\beta}_{g_{i},j}^{0}(\lambda_{g_{i}})\neq 0. \,\,\, j=1,\ldots,p\}.
\end{equation}

For data analysis, we created training ($T_{r_{g_{i}}}$) and test  ($T_{s_{g_{i}}}$) sets of equal size via simple random sampling. The train set ($T_{r_{g_{i}}}$) is taken under study and we separated it into G groups $(X_{1},X_{2},\ldots,X_{G})$. With replacement samples $\left(X_{1}^{*},X_{2}^{*},\ldots,X_{G}^{*}\right)$ are taken from each group $X_{G}$, where $X_{g_{i}}^{*}$ contains hundred with replacement samples $\Big(\left((x_{g_{i},1}^{*})\right),\left((x_{g_{i},2}^{*})\right),\ldots,\left((x_{g_{i},100}^{*})\right) \Big)$. As we know the LASSO is a convex optimization problem which uses $\ell_{1}$-penalty and solves the following problem
\begin{equation}
    \label{lasso}
       \hat{\beta}_{lasso}(\lambda) = \arg\min_{\beta}{\sum_{i}(y_{i}-x_{i}^{T}\beta)^{2} + \lambda \|\beta\|_{1} },
\end{equation}
where $\lambda$ is usual regularization parameter which is chosen via 10 fold-CV. We then implemented the LASSO algorithm (\ref{lasso}) over the bootstrapped samples $\big(\left(x_{g_{i},k}^{*}\right)\big)$ where $1\leq k \leq 100$. This gives us hundred different set of active set of variables $\hat{S}_{g_{i},k}^{*}(\lambda_{g_{i},k}^{*})$.  The union of the set $\hat{S}_{g_{i},k}^{*}(\lambda_{g_{i},k}^{*})$ resulted in a lower dimensional dataset $X_{g_{i}}^{**}$. The new designed matrix $X_{g_{i}}^{**}$ can be expressed as;
\begin{equation}
    X_{g_{i}}^{**} = \bigcup_{k=1}^{100}S_{g_{i},k}^{*}(\lambda_{g_{i},k}^{*}).
\end{equation}

The above procedure is repeated for other groups over the bootstrapped sample $(X_{1}^{*},X_{2}^{*},\ldots,X_{G}^{*})$ and we get different sets of active set variables that results in different lower dimensional datasets $\big(X_{1}^{**},X_{2}^{**},\ldots,X_{G}^{**}\big)$.

In addition to this approach we performed another bootstrapped resampling over the dataset $X$ called bootstrapped DSL. For our purpose primarily we divide the dataset into two halves, and then we separate the dataset group wise and take with replacement samples from each group  $\big(X_{1}^{*},X_{2}^{*},\ldots,X_{G}^{*}\big)$ each of size hundred from the train set $T_{r{g{i}}}$. Where 
\begin{center}
$X_{1}^{*} = \Big(\big(\left(x_{1,1}^{**}\right)\big),\big(\left(x_{1,2}^{**}\right)\big),\ldots,\big(\left(x_{1,100}^{**}\right)\big)\Big)$,\\
 $ X_{2}^{*}= \Big(\big(\left(x_{2,1}^{**}\right)\big),\big(\left(x_{2,2}^{**}\right)\big),\ldots,\big(\left(x_{2,100}^{**}\right)\big)\Big)$, \\ \ldots, \\
 $ X_{G}^{*} = \Big(\big(\left(x_{G,1}^{**}\right)\big),\big(\left(x_{G,2}^{**}\right)\big),\ldots,\big(\left(x_{G,100}^{**}\right)\big)\Big).$
\end{center}
To implement the DSL algorithm we prepared the data matrices $Z_{k}^{*}$ (where $1\leq k \leq 100$) as defined in \cite{gross}, with the help of the bootstrapped samples $\big(X_{1}^{*},X_{2}^{*},\ldots,X_{G}^{*}\big)$, as $\big(Z_{1}^{*},Z_{2}^{*},\ldots,Z_{100}^{*}\big).$ The DSL algorithm is implemented over each bootstrapped sample $\big(Z_{k}^{*}\big)$ which gives us different set of active set variables
$\hat{S}_{z}^{*}(\lambda^{*})$ = $\left(\hat{S}_{z_{1}}^{*}(\lambda_{1}^{*}), \hat{S}_{z_{2}}^{*}(\lambda_{2}^{*}),\ldots,\hat{S}_{z_{100}}^{*}(\lambda_{100}^{*})\right)$.
The union of the set $\hat{S}_{z}^{*}(\lambda^{*})$, resulted in a dataset $Z^{**}$ having lower dimension which can be expressed as
\begin{equation}
    Z^{**} = \bigcup_{k=1}^{100}\hat{S}_{z_{i}}^{*}(\lambda_{z_{i}}^{*}),
\end{equation}

to check the significance of proposed resampling procedures, we implement LASSO and DSL algorithms on the resulting data sets. The results are explained in section (\ref{results}).

\subsection{Improvisation in Data Shared Lasso Model}
\label{noise}
\begin{figure}[htb!]
\centering
 \includegraphics[width=160mm]{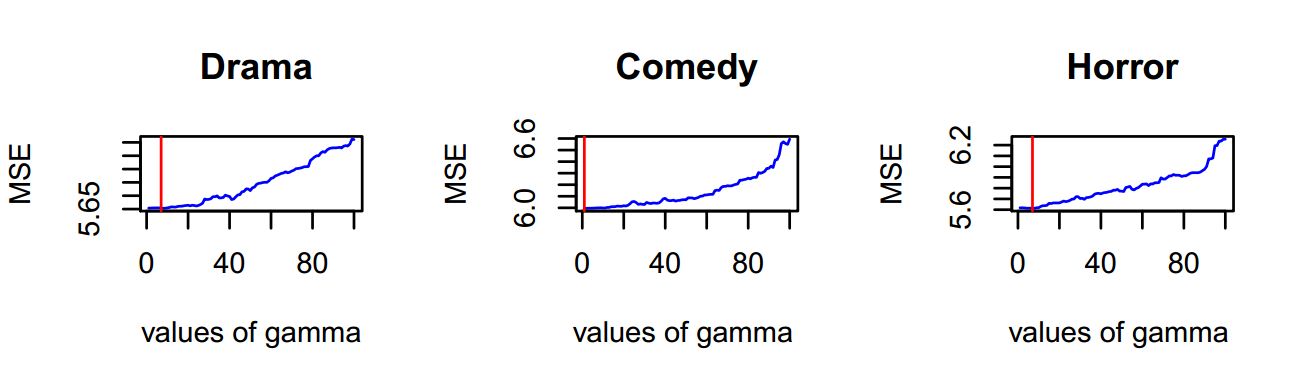}
 \caption{These plots are showing the variability in the mean squared by varying the value of $\gamma_{1}$ in Donoho's global de-noising (\ref{donoho}) threshold. Here x-axis is showing the values of $\gamma_{1}$, a vector of size hundred ($\gamma_{1}$ $\in$ $[0,1/2 ]$) and the relative change in MSE is in y-axis. The red vertical line is showing the point at which the minimum of the mean squared error occurred.}
 \label{fig:donoho}
\end{figure}

Choice of the regularization parameter $r_{g}$ in DSL model plays an important role in the sense that it controls the amount of sharing  within the groups. The DSL model can be seen as LASSO model \cite{gross}, with the following representation:
\begin{equation}
\label{dsllasso}
   \frac{1}{2}\|\tilde{y}-Z\tilde{\beta}\|^{2} + \lambda \|\tilde{\beta}\|_{1}  = \frac{1}{2}\sum_{i} \left(y_{i}-x_{i}^{T}(\beta + \Delta_{g_{i}})\right)^{2} + \lambda \left(\|\beta\|_{1} + \sum_{g=1}^{G}r_{g}\|\Delta_{g}\|_{1}\right).
\end{equation}
The above representation allows to apply LASSO algorithm directly to solve the DSL optimization problem. We apply de-noising threshold on the selected LASSO model from equation (\ref{dsllasso}). By varying the value of $\gamma_{1}$ $\in$ $[0,1/2]$ in (\ref{donoho}), we observed the variability in the mean squared error. A pictorial representation of the variability in MSE by varying the constant $\gamma_{1}$ is shown through figure (\ref{fig:donoho}).

From the following figure (\ref{fig:donoho}) we see that by varying the value of $\gamma_{1}$ $\in$ $[0,1/2]$, after a certain point the MSE is increasing rapidly and the minimum of the mean squared error is attained at a different point other than $(\gamma_{1} = 0)$, which indicates the presence of noise in the extracted features.

The objective function (\ref{DSL}) is overparameterized as typically cardinality of $\beta$ and $\Delta_{g}$ is larger than sample size $n$. Therefore we need to penalize the $\Delta_{g}$ term by $\ell_{1}$ penalty to reduce the effective number of parameters. Following \cite{gross}, for $r_{g}=r$, the penalty term simplifies to,
\begin{equation*}
  \lambda (\|\beta\|_{1} + \sum_{g}r_{g} \|\Delta_{g}\|_{1}) = \lambda (\|\beta\|_{1} + \sum_{g}r \|\Delta_{g}\|_{1}) 
\end{equation*}

\begin{equation*}
    = \lambda r (\|\beta\|_{1}/r + \sum_{g} \|\Delta_{g}\|_{1} = \lambda^{new} (\|\beta\|_{1}/r + \sum_{g} \|\Delta_{g}\|_{1})
\end{equation*}

where $\lambda^{new}$ = $\lambda r$. 

Let us consider the case when we have unequal number of sample sizes i.e. $n_{g_{1}} \neq n_{g_{2}} \neq \cdots \neq n_{G}$, where $n_{g_{i}}$ denote the sample size of each group. Certain arguments have been given in \cite{gross} for choosing the value of regularization parameter $r_{g}$.

Following \cite{gross} has shown that if we assume that we have fixed $\beta_{g}^{*}$ $\in$ $\mathcal{R}^{p}$ and $r_{g}$ such that $\beta_{g}^{*} = \beta + \Delta_{g}$ then the problem
\begin{equation}
   \label{penalty}
    (\hat{\beta},\hat{\Delta}_{g})= \lambda (\|\beta\|_{1} + \sum_{g}r_{g} \|\Delta_{g}\|_{1}),
\end{equation}

is separable in $p$, for the case $p=1$, all values that satisfy the equality constraint fall into a family that can be characterized as $\beta = c$, and $\Delta_{g} = \beta_{g}^{*} - c$ hence (\ref{penalty}) is equivalent to solving unconstrained optimization in one variable: $\hat{c} = \arg\min|c| + \sum_{g}r_{g}|\beta_{g}^{*}-c|$ and the constraint $\sum_{g}{r_{g}} < 1$ leads to separate regression. From the above discussion it is clear that the value of $r_{g}$ would be chosen under the condition $\sum_{g}{r_{g}} > 1$. When we have unequal number of observations the above problem (\ref{penalty}) can be simplified as:
\begin{center}
    $\lambda (\|\beta\|_{1} + \sum_{g}r_{g} \|\Delta_{g}\|_{1})= \lambda (\|\beta\|_{1} + r_{1} \|\Delta_{1}\|_{1} + r_{2} \|\Delta_{2}\|_{1},\cdots,r_{G} \|\Delta_{G}\|_{1})$\\
   \hspace{4.4cm}     $=\lambda \|\beta\|_{1} + \lambda(r_{1} \|\Delta_{1}\|_{1} + r_{2} \|\Delta_{2}\|_{1},\cdots,r_{G} \|\Delta_{G}\|_{1})$\\
   \hspace{4.0cm}     $=\lambda \|\beta\|_{1} + r_{1}^{*} \|\Delta_{1}\|_{1} + r_{2}^{*} \|\Delta_{2}\|_{1},\cdots,r_{G}^{*} \|\Delta_{G}\|_{1},$
\end{center}
where $r_{g_{i}}^{*} = \lambda r_{g_{i}}$ are new regularization parameter over the groups. As we have unequal number of observation so the chosen penalty should depend on the sample size under the condition $\sum_{g}{r_{g}} > 1$.  Suggested regularization parameters are chosen with the help of universal global de-noising threshold (\ref{donoho}) that reduces the noise from the extracted features.
The penalties used in the DSL model are following $\sqrt{\frac{1}{3}}$, $\sqrt{\frac{n_{g_{i}}}{N}} $,$\sqrt{\frac{\log{N}}{\log{n_{i}}}} $,$\frac{N}{n_{g{i}}}$ ,$\frac{\log{N}}{\log{n_{g{i}}}} $, $\frac{\log{n_{g{i}}}}{\log{N}}$ and $\sqrt{\frac{\log{n_{g_{i}}}\times N}{\log{N}\times n_{g_{i}}}}$ respectively over different groups, where $N = n_{g_{1}}+n_{g_{2}}+n_{g_{3}}$.\\

\subsection{A Combined Study of DSL and Separate LASSO Together With Differential Regularization Scheme}
\label{new_noise}
Another attempt to reduce the noise from the extracted features by using both DSL and LASSO together, proposed in this article. 
For our purpose firstly we implement LASSO algorithm over  groups ($X_{1},X_{2}, \ldots, X_{G}$) separately which gives us different set of active set variables $\hat{S}_{g_{i}}^{*}(\lambda_{g_{i}}^{*})$ where $1\leq g_{i}\leq G$. Next we implemented the DSL algorithm over the matrix $Z$ together with differential regularization which gives us another different set of active set variables $\hat{S}_{z,(r_{g})_{j}}^{*}(\lambda_{(r_{g})_{j}}^{*})$. Here we are interested in shared active set variables in the set $\hat{S}_{z,(r_{g})_{j}}^{*}(\lambda_{(r_{g})_{j}}^{*})$ and $\hat{S}_{g_{i}}^{*}(\lambda_{g_{i}}^{*})$ for $1\leq g_{i}\leq G$. Let the set of shared active set variables be denoted as $\hat{S}_{z_{s},(r_{g})_{j}}^{*}(\lambda_{(r_{g})_{j}}^{*})$. The main idea is to create different subgroups by using both set of active set variables $\hat{S}_{g_{i}}^{*}(\lambda_{g_{i}}^{*})$ and $\hat{S}_{z_{s},(r_{g})_{j}}^{*}(\lambda_{(r_{g})_{j}}^{*})$ and study the relative change in MSE by removing different subgroups from the main model. These subgroups are created by taking the intersection between different groups which are as follows
\begin{center}
$\left(\hat{S}_{g_{i}}^{*}(\lambda_{g_{i}}^{*})\right)\bigcap \left(\hat{S}_{z_{s},(r_{g})_{j}}^{*}(\lambda_{(r_{g})_{j}}^{*})\right),$\\
$\left(\bigcap_{g_{i}=1}^{G}\hat{S}_{g_{i}}^{*}(\lambda_{g_{i}}^{*})\right)$ $\bigcap$ $\left(\hat{S}_{z_{s},(r_{g})_{j}}^{*}(\lambda_{(r_{g})_{j}}^{*})\right)$,\\
$\left(\hat{S}_{z_{s},(r_{g})_{j}}^{*}(\lambda_{(r_{g})_{j}}^{*})\right)- \left(\bigcup_{g_{i}}^{G} \hat{S}_{g_{i}}^{*}(\lambda_{g_{i}}^{*})\right)$,
\end{center}

where $g_{i} (= 1,2,...,G)$ denote the number of groups and $j (= 1,2,...,l)$ denote the number of penalty used. The above proposed technique help us to classify all those set of features which improves the prediction accuracy.

\begin{figure}
\centering
\subfloat[Drama]{
  \includegraphics[width=70mm]{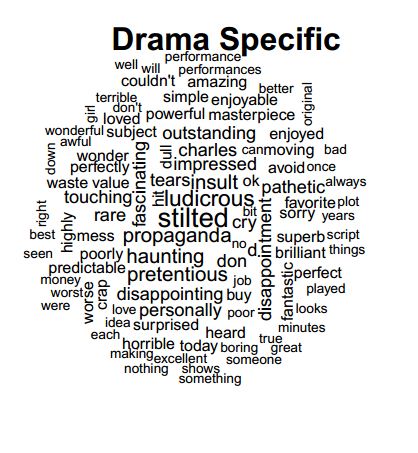}
}
\subfloat[Comedy]{
  \includegraphics[width=70mm]{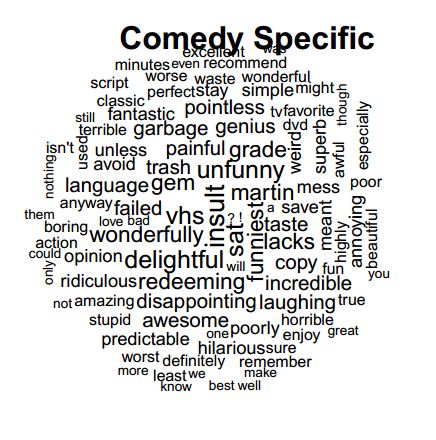}
}

\subfloat[Horror]{   
  \includegraphics[width=70mm]{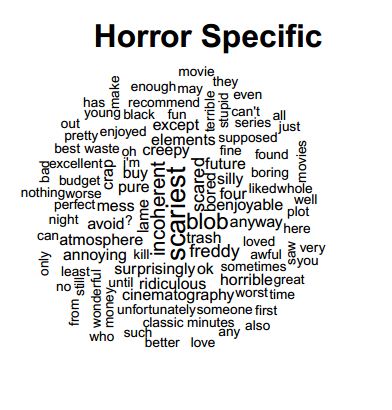}
}
\caption{This word cloud is for the coefficients found after running Bootstrapped LASSO model separately on each genre. The size of the words are indicating proportion of words appeared in 100 bootstrap sample}
\end{figure}

\section{ Experimental Setup for IMDb Data}
For our purpose we used publicly available IMDb dataset review dataset aclImdb \cite{mass} of movie reviews from ~\url{IMDb.com}. which contains set of reviews and corresponding ratings associated with the them where the dependent variable $Y_{i}$'s are integer rating and explanatory variable $X_{i}$'s are text features. The text features are converted into integers by taking the number of occurrences of each feature in each review. This resulted in a high dimensional dataset that is the sample size is less than the features. Later the entries of the predictor matrix $X$ are converted into binary matrix and the sparse representation of the data matrix $X$ is used for the analysis that is
\begin{equation}
  x_{ij}=
    \begin{cases}
    1, & \text{if $j^{th}$ feature is present in $i^{th}$ review }\\
    0, & \text{otherwise}.
    \end{cases}
\end{equation}

As there are $2^{p}$ possible sub-models, computational feasibility is crucial and when the data passes through the LASSO \cite{tibshirani} it selects at most n variables before it saturates. To improve the computation feasibility of the model we need to compress the size of the bag of features.

Let us assume that $X_{G}$ come from  $N_p(0,\Sigma)$, where $\Sigma \succ 0.$ We divide the dataset $X_{G}$ into two halves, train $T_{r_{g_{i}}}$ and test $T_{s_{g_{i}}}$ set. The DSL algorithm is formulated for the supervised learning problem which is applied to a sentiment analysis dataset. This dataset contains 50k movie reviews, allowing no more than 30 reviews per movie that have been divided into two sets of equal size training and test sets. This dataset contains polarized reviews half of the reviews are positive (rating $\geq7$) and other half are negative (rating $\leq4$). Binary \textit{bag-of-words} representation of the reviews is used. Only those words that appear in at least five reviews in our training set, are considered for the analysis of the dataset. This resulted $p = 27743$ features in our dataset. Response value is integer rating. To implement the DSL algorithm, we mainly focused on genres of the movies. The DSL model is fitted over three most popular genres drama, comedy and horror. Only those review are considered which appear in at least one of the three genres. This resulted in $n = 16386$ reviews in training set (8286 dramas, 5027 comedies, 3073 horror movies). The DSL algorithm can be implemented with any lasso solver using a straightforward augmented data approach:

Following \cite{gross} defined the data matrix Z as
\\
\begin{center}
\(
Z=
\begin{bmatrix}
X_{1} & r_{1}X_{1} & 0 & .&.&.&0\\
X_{2} & 0 & r_{2}X_{2} &.&.&.&0\\
.\\
.\\
.\\
X_{G} & 0 & 0 &.&.&.&r_{G}X_{G}
\end{bmatrix}
\),

\end{center}

where $X_{j}$ and $y_{j}$  denote the data matrix for the group $g_{i}=j$. The above problem is solved by R package \textbf{glmnet} \cite{friedman} which has capability to use sparse representation of a data matrix. It is quite faster than other LASSO solver.
Here it is easy to change the family and loss function. In \textbf{glmnet} package we can fit ridge as well as elastic net by changing the values of $\alpha$. For ($\alpha = 1$) glmnet fits LASSO model, for $(\alpha = 0)$ it fits ridge regression and for $\alpha \in (0,1)$ it fits elastic net model.



\section{Experimental Result}
\label{results}

\begin{figure}[htbp!]
\centering
\includegraphics[width=160mm,height=55mm]{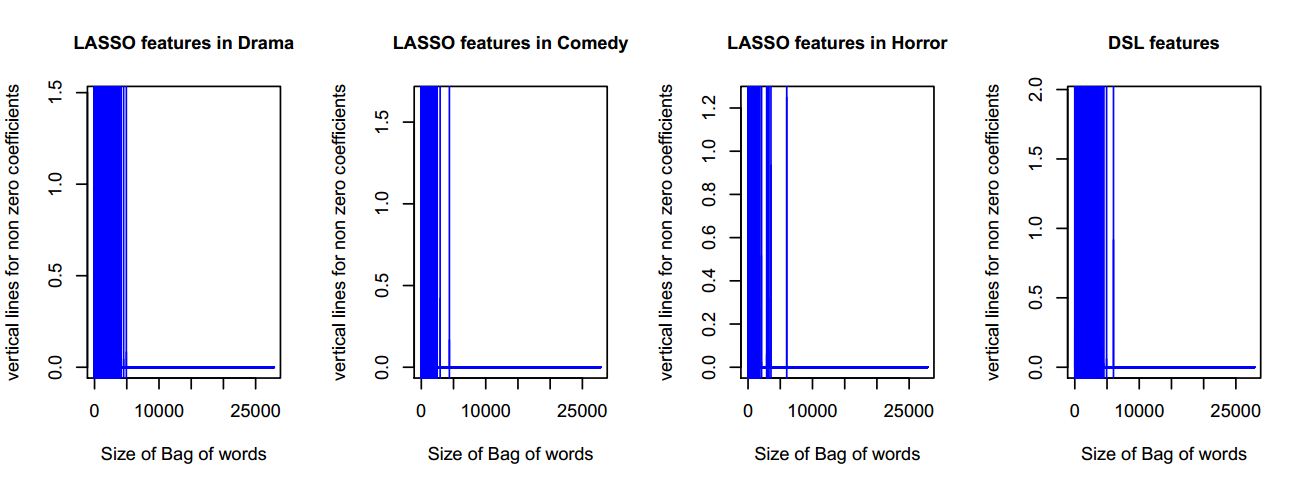}
\caption{Here the vertical lines are showing the presence of the non zero coefficients after running LASSO in each genre and DSL on $Z$.}
\label{fig:bagofwords_comp1}
\end{figure}

\begin{figure}[htbp!]
\centering
\includegraphics[width=160mm,height=55mm]{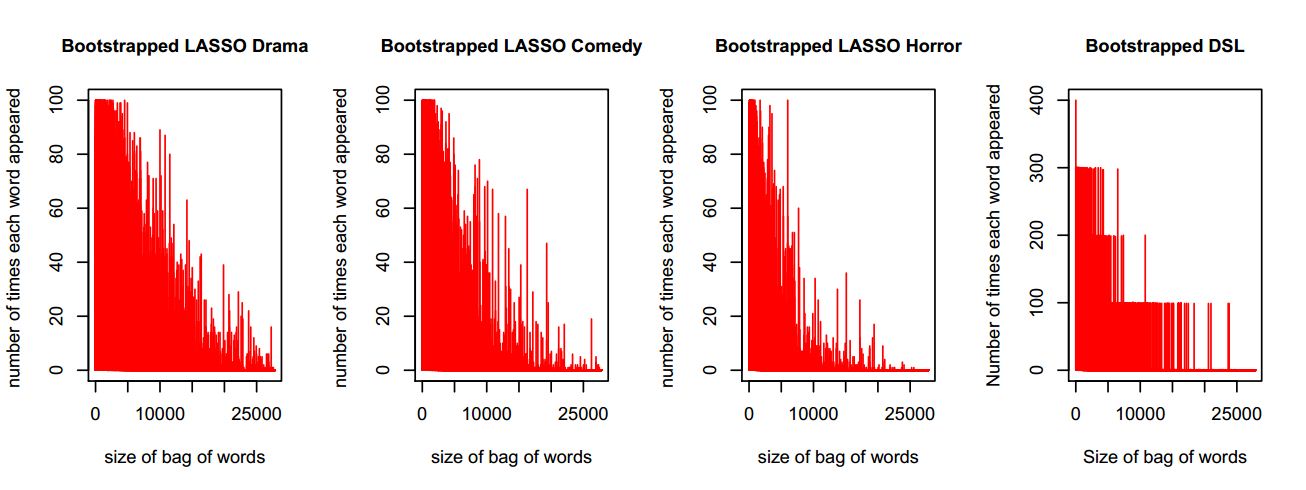}
\caption{Here are the number of times each word found from running bootstrapped LASSO on each genre and bootstrapped DSL on matrix $Z$. Here x-axis is showing the size of the bag of words and y axis is showing number of times each word appeared.}
\label{fig:bagofwords_comp}
\end{figure}

We implemented our methodology to IMDb dataset, we performed the data analysis for the various choice of the group regularization parameter $r_{g}$. 

Following fig.(\ref{fig:bagofwords_comp1}) shows the nonzero coefficients appeared in the main model. Blue vertical lines are showing the presence of nonzero coefficient in both the main model. First three pictures are showing for LASSO and fourth is showing for DSL model. Following fig.(\ref{fig:bagofwords_comp1}) showing the results obtained after fitting the Bootstrapped LASSO and Bootstrapped DSL. Here each horizontal red line is showing the number of times a coefficient appeared in hundred bootstrap sample. First three pictures in fig.(\ref{fig:bagofwords_comp1}) are obtained after running bootstrapped LASSO in each genre and the fourth picture is showing the results obtained after running bootstrapped DSL.

From both the figure (\ref{fig:bagofwords_comp1}) and (\ref{fig:bagofwords_comp}) we see that there are many features which appeared more than sixty times while they did not appear in the main model. This is the one of the advantage of our proposed technique that it gives us all the important sentiments which are in the dataset but not selected in the main model. These sentiments can be extracted by the proper choice of the threshold over the fraction of the extracted features. Another advantage of this technique is that the union of the all the extracted features resulted in a lower dimensional dataset hence it can be think of as a one of the dimension reduction technique. This reduces the size of the \textit{bag-of-features} for drama it reduces to 9854 from 27743 for comedy it reduces to 6201 and for horror it reduces to 4114 features from 27743 features and also the column size of the entries of the matrix $Z$ \cite{gross}, also reduces to 11295 from 27743 features.

Proposed methodology gives us two types of dataset of different dimension. We performed data analysis over both the dataset the results are given in table (\ref{table:1}) and (\ref{table:2}). 
We run LASSO in each genre and DSL on $Z$ on both type of dataset and results are shown in table (\ref{table:1}) and (\ref{table:2}). Here we can see that the prediction accuracy of both the data set after running LASSO and DSL model is same as main dataset. These dataset fastens the time required for performing same data analysis and removes the redundant features.

\begin{table}[htbp]
\caption{Comparison of test set mean squared error after running LASSO in each genre of both type of dataset with main dataset. }
\vskip10pt
\centering
\begin{tabular}{|c|c|c|c|}
    \hline
    &Drama&Comedy&Horror  \\
    \hline
   Old data &  5.66 &  5.99 &   5.63 \\
   \hline
    BLS  &  5.66 &  5.99 &  5.63 \\
    \hline
    BSLS  &  5.66 &  5.99 &  5.63 \\
    \hline
\end{tabular}
\label{table:1}
\end{table}

\begin{table}[htbp]
\caption{Comparison of test set mean squared error after running DSL in both type of dataset with main dataset.}
\vskip10pt
\centering
\begin{tabular}{|c|c|c|c|c|c|}
     \hline\hline
     &Model Type &All&Drama&Comedy&Horror  \\
     \hline\hline
    Old Data &Pooled&  5.63 & 5.57 &  \textbf{5.78} &   5.55 \\
    BLS &Pooled&  5.63 & 5.57 &  \textbf{5.78} &   5.55\\
    BSLS &Pooled &5.62 & \textbf{5.55} &  \textbf{5.78} &   5.57\\
    \hline\hline
    Old Data & separate &  5.72 &  5.63 & 5.97 &   5.61 \\
    BLS & separate &  5.72 &  5.63 & 5.97 &   5.61 \\
    BSLS& separate & 5.70 & 5.61 & 5.95 &   5.59 \\
    \hline\hline
    Old Data & Data Shared & \textbf{5.54} &  \textbf{5.56} & 5.85 & \textbf{5.07} \\
    BLS & Data Shared & \textbf{5.54} &  \textbf{5.56} & 5.85 & \textbf{5.07} \\
    BSLS & Data Shared& \textbf{5.54} &  5.56 & 5.85 & \textbf{ 5.07} \\
     \hline\hline
\end{tabular} 
\label{table:2}
\end{table} 

Comparison of extracted features through venn diagram is shown in fig.(\ref{fig:6}). Following fig.(\ref{fig:6}) is showing venn diagram comparison of the features extracted after running DSL model in all three types of dataset. The sub fig.(a) is for the dataset with 27743 features, sub fig.(b) is for the dataset with 11295 features  and sub fig.(c) is for the dataset with 4499 features. From the venn diagram representation we can see that feature distribution is almost same in all three types of the dataset.

\begin{figure}
\centering
\subfloat[Main Data]{
  \includegraphics[width=65mm]{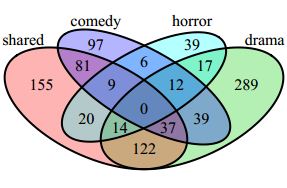}
}
\subfloat[BLS]{
  \includegraphics[width=65mm]{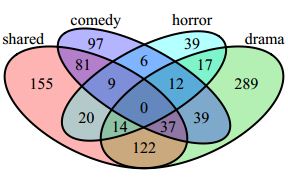}
}
\hspace{0mm}
\subfloat[BSLS]{
  \includegraphics[width=65mm]{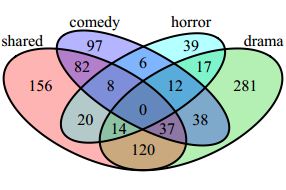}
}
\caption{The comparison of the extracted feature after running DSL model over all types of dataset.}
\label{fig:6}
\end{figure}

We implemented our methodology (\ref{noise}) on IMDb dataset. Here we evaluate DSL model with differential regularization and the results are given in following table (\ref{table:3}). From the table (\ref{table:3}) we see that the average mean squared difference between true rating and predicted rating varies with the choice of the regularization parameter $r_{g}$. Selected regularization parameter $r_{g}$ performs better in terms of MSE.

\begin{table}[htbp]
\caption{Comparison of test set MSE using different weighting scheme over groups.}
\vskip10pt
\centering
\begin{tabular}{|c|c|c|c|c|c|}
    \hline
   Penalty type&Model&All&Drama&Comedy&Horror  \\
   \hline\hline
   $\sqrt{\frac{1}{3}}$ &Data Shared & 5.54 &  5.56 & 5.85 &  5.07 \\
    \hline\hline
   $\sqrt{\frac{n_{g_{i}}}{N}}$ &Data Shared & 5.57 &  5.63 &  5.84 &  5.05 \\
   \hline\hline
   $\sqrt{\frac{\log{N}}{\log{n_{g_{i}}}}}$  &Data Shared & 5.43 &  5.47 &  5.71 &  4.93 \\
    \hline\hline
    $\frac{N}{n_{g{i}}}$&Data shared &5.42 & 5.43 & 5.71 & 4.99\\
    \hline\hline
    $\frac{\log{N}}{\log{n_{g{i}}}} $&Data shared &5.43& 5.47 &5.71&4.93\\
    \hline\hline
   $\frac{\log{n_{g{i}}}}{\log{N}}$ &Data shared& 5.46 & 5.49 & 5.74 & 4.97\\
    \hline\hline
    $\sqrt{\frac{\log{n_{g_{i}}}}{\log{N}}}$&Data shared&5.46&5.50&5.74&4.95\\
    \hline\hline
    $\sqrt{\frac{\log{n_{g_{i}}}\times N}{\log{N}\times n_{g_{i}}}}$&Data shared&5.42&5.44&5.69&4.96\\
    \hline\hline
   
\end{tabular} 
\label{table:3}
\end{table}

\begin{figure}[htbp!]
\centering
\subfloat[$\sqrt{\frac{1}{3}}$]{
  \includegraphics[width=60mm]{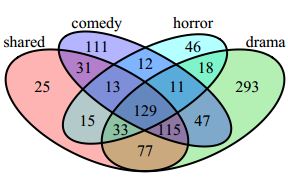}
}
\subfloat[$\sqrt{\frac{n_{g_{i}}}{N}}$]{
  \includegraphics[width=60mm]{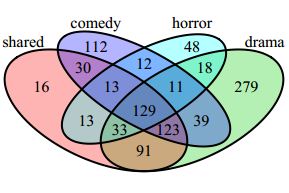}
}
\hspace{0mm}
\subfloat[$\sqrt{\frac{\log{n_{g_{i}}}}{\log{N}}}$]{
  \includegraphics[width=60mm]{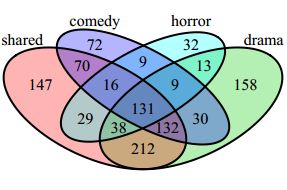}
}
\subfloat[$\sqrt{\frac{\log{N}}{\log{n_{g_{i}}}}}$]{
  \includegraphics[width=60mm]{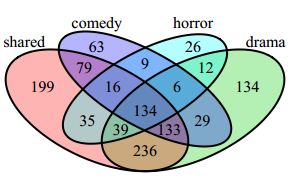}
}
\hspace{0mm}
\subfloat[$\frac{\log{n_{g_{i}}}}{\log{N}}$]{   
  \includegraphics[width=60mm]{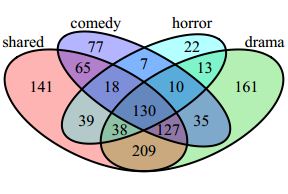}
}
\subfloat[$\frac{\log{N}}{\log{n_{g_{i}}}} $]{
  \includegraphics[width=60mm]{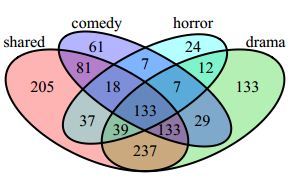}
}
\hspace{0mm}
\subfloat[$\sqrt{\frac{\log{n_{g_{i}}}\times N}{\log{N}\times n_{g_{i}}}}$]{   
  \includegraphics[width=60mm]{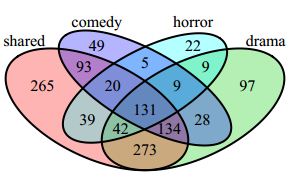}
}
\caption[Short caption]{Here are the venn diagrams between the coefficient found after running LASSO on each genre and shared coefficient found after running DSL model for different values of regularization parameter $r_{g}$  on an IMDb sentiment analysis dataset.}
\label{fig:8venn}
\end{figure}

Following table (\ref{table:6}) is showing the relative change in test set MSE in percentage in view of fig.(\ref{fig:8venn}). Here we study the effect of different subgroups on DSL model with differential regularization $r_{g}$. We implemented our methodology (\ref{new_noise}) on IMDb dataset. Following fig.(\ref{fig:8venn}) showing the venn diagram representation of the shared coefficient and separate coefficients for different regularization. Each diagram in fig.(\ref{fig:8venn}) is showing results obtained after imposing a different penalty over groups. Here the size of each ellipse is proportional to the number of features in that ellipse. The diagrams in fig.(\ref{fig:8venn}) are divided in different subgroup of features. Following table (\ref{table:6}) is showing the relative effect of different subgroups in terms of mean squared error. Here we removed different subgroups which are in common with shared features and the relative effect in percentage is noted in table (\ref{table:6}). From the fig.(\ref{fig:8venn}) we see that there are 130 features which are in common for every choice of the regularization parameter $r_{g}$ and also the removal of these words increases the relative MSE around $23\%$ which explains the importance of these features in the model.
From the table (\ref{table:6}) we see that there is no significant relative change in MSE after removing the additional features which are added to shared features which leads to noise in the sub model.

\section{Discussion}

In this article we present a new approach to choose reduce the size of \textit{bag-of-words} and also an idea about choosing the regularization parameter $r_{g}$. The generalized data enriched model \ref{DSLopt} can be written as
\begin{center}
$y_{i} = x_{i}^{T}\beta + x_{i}^{T}C_{g_{i}}\Delta_{g_{i}}/C_{g_{i}} + \epsilon_{i}$,
\end{center}
where $C_{g_{i}}$ is the variability order of the normalization column $X_{i}$. This can be written as
\begin{center}
$y_{i} = x_{i}^{T}\beta + x_{i}^{T}\gamma_{g_{i}}/C_{g_{i}} + \epsilon_{i}$,
\end{center}
where $\gamma_{g_{i}}$ = $C_{g_{i}}\Delta_{g_{i}}$ are new group effects. The penalty term simplifies to,
\begin{center}
$\lambda(C_{0}\|\beta\|_{1} + \sum_{g=1}^{G}r_{g}C_{g}\Delta_{g})$=
$\lambda(C_{0}\|\beta\|_{1} + \sum_{g=1}^{G}w_{g}\Delta_{g}$),
\end{center}
where $w_{g}$ = $r_{g}C_{g}$ are new weights over the groups.

Standardization of columns should be done on different groups and pooled groups. Under normal thresholding when $\mathcal{O}(\sqrt{n}\hat{\beta})$ are $\mathcal{O}_{P}(1)$ in such of case Donho and Jhonstone find for declaring global threshold (\ref{donoho}) is of order $\sqrt{2log(n)}.\gamma_{1}.\sigma/\sqrt{n}$. Since we have sparse matrices so order may be little low of n and since order is not known we tried different choice of n. From the equation (\ref{donoho}) we have
\begin{center}
$t_{n}=\sqrt{2log(n)}.\gamma_{1}.\sigma/\sqrt{n}$,
\end{center}
this can be written as 
\begin{center}
$\sqrt{\frac{n}{log(n)}}=\sqrt{2}\gamma_{1}.\sigma/t_{n} = r_{g}$.
\end{center}

Different choice of $C_{0}$ gives different choice for the regularization parameter $r_{g}$. \\
Let $N=\sum_{g_{i}=1}^{G}n_{g_{i}}$ then we have\\
1) Choosing $n=n_{g_{i}} = n$, $log(n)=log(N)$ and $C_{0}$ = $N/log(N)$ weights over groups becomes $\sqrt{1/3}$.\\
2) Choosing $n=n_{g_{i}}$, $log(n)=log(N)$ and $C_{0}$ = $N/log(N)$ weights over groups becomes $\sqrt{(n_{g_{i}}/N)}$.\\
3) Choosing $n=log(n_{g_{i}})$, $log(n)=n_{g_{i}}$, $C_{0}$ = $\sqrt{log(N)/N}$ weights over groups becomes $\sqrt{\frac{\log{n_{g_{i}}}\times N}{\log{N}\times n_{g_{i}}}}$.\\
4) Choosing $n=\log{n_{g_{i}}}$, $log(n)=log(N)$ and $C_{0}$ =1 weights over groups becomes $\sqrt{\log{n_{g_{i}}}/\log{N}}$.\\
5) Choosing $n=log(N)$, $log(n)=\log{n_{g_{i}}}$ and $C_{0}$ =1 weights over groups becomes $\sqrt{\log{N}/\log{n_{g_{i}}}}$.

By taking square of the penalties $(4)$ and $(5)$ we can find other two penalties. Proposed weights satisfy the condition $\sum_{g}{r_{g}} > 1$.

Bootstrapped LASSO and DSL can be think of a data reduction technique which reduces the size of \textit{bag-of-features} that keeps all the important features available in the dataset. Fig. $(2)$ is showing the word cloud for the top hundred sentiments in terms of proportion of words appeared after running Bootstrapped LASSO on each genre. The table (\ref{table:3}) shows that the choice of the differential regularization also has improved prediction accuracy in terms of MSE. From the table (\ref{table:6}) we see that additional features added to shared features can be assumed to be noise as there is no greater change in relative MSE. Those features which are common in different groups seems to be important sentiments as removal of these features results in the greater change in relative MSE (\ref{table:6}) and also fig. (\ref{fig:8venn}) depicts that there are almost 130 features which are common in all groups as well in shared features for all types of penalty and removal of these features has greater change in relative MSE which indicates that this representation gives us smallest subset of the important sentiments.

\section*{Acknowledgement}
The author would like to thank Professor Debapriya Sengupta, Applied Statistics Unit, Indian Statistical Institute, Kolkata, for his constructive collaboration in this work.

\begin{table}[htbp!]
 \caption[Short caption]{Following table shows the relative  change in MSE in percentage by removing different subgroup of features in view of figure (\ref{fig:8venn}). Following table explains the importance of the subgroups of the features based on relative change in MSE. The relative change in MSE is noted after removing the no of features (which are in right most column) from the main model.}
 \vskip4pt
 \centering
 \scalebox{0.8}{
 \begin{tabular}{|c|c|c|c|c|c|c|}
    \hline
    &&MSE &MSE &MSE &MSE &\\
   Penalty &Removal type&$\%$ increase &$\%$ increase&$\%$ increase&$\%$ increase  &No.of coef. \\
   type&&in All&in Drama&in Comedy&in Horror&removed\\
   \hline\hline
   $\sqrt{\frac{1}{3}}$&no removal & 0 &  0 & 0 & 0&0 \\
   &all intersection  &25.8 &23.5 & 31.6 & 23.0 & 129\\
   &shared int dram  &40.2 &39.0 &46.4& 33.6 &354\\
   &shared int com  &36.7 &34.7 &43.5 & 31.6 &288\\
   &shared int hor  &28.4&25.9 &35.2 & 25.1 &190\\
   &additional &0.4 &0.5 &0.6 & 0.1&25 \\
   \hline\hline
   $\sqrt{\frac{n_{g_{i}}}{N}}$&no removal &0 &  0 &  0 & 0 &0 \\
   &all intersection  &24.5 &22.3 & 29.9 & 22.1 &129 \\
   &shared int dram  &40.1 &39.5 &45.0 & 33.9 &376\\
   &shared int com  &36.1 &34.0 &42.6 & 31.3 &295\\
   &shared int hor  & 27.2 & 24.9 &33.6&  23.6 &188\\
   &additional &0.04 & 0.02 &0.05 & 0.02 &16\\
   \hline\hline
   $\sqrt{\frac{\log{N}}{\log{n_{g_{i}}}}}$&no removal&0&0&0 &0 &0\\
   &all intersection&23.9 &22.3 & 28.7 & 20.7 &134 \\
   &shared int dram  &42.4 &43.8 & 45.7 & 33.1 &542\\
   &shared int com & 37.9 &35.9 &45.5 & 30.9 &362\\
   &shared int hor  & 26.8&25.3&32.4&  21.8 &224\\
   &additional &1.1 &1.3 &1.2&0.7 &199\\
   \hline\hline
   $\sqrt{\frac{\log{n_{g_{i}}}}{\log{N}}}$&no removal &0&0&0& 0& 0\\
   &all intersection  &23.6 &21.9 & 28.3 & 20.8 & 31 \\
   &shared int dram &42.0 & 43.2 &45.9 &  32.1 &513\\
   &shared int com & 37.0 &35.4 &44.1 & 30.5 &349\\
   &shared int hor  & 26.4&24.6&32.0& 21.9 &214\\
   &additional &1.0 &1.2 & 1.1 & 0.6 &147\\
   \hline\hline
   
   $\frac{\log{n_{g{i}}}}{\log{N}}$&no removal  & 0 & 0 & 0 & 0&0\\
   &all intersection  &23.2  &21.9 & 27.5 & 20.1&130 \\
   &shared int dram &41.2 & 42.7 &44.5 & 31.6 &504\\
   &shared int com  & 36.9 &35.9 &43.1 & 29.8 &340\\
   &shared int hor & 26.2 &24.8&31.3& 22.1 &225\\
   &additional &1.1 & 1.2 & 1.1 & 0.7 &141\\
   \hline\hline
   
   $\frac{\log{N}}{\log{n_{g{i}}}} $&no removal &0 &0 & 0 &0 & 0\\
   &all intersection  &23.4 &21.8 & 28.1 & 20.6 &133 \\
   &shared int dram &42.1 & 43.6 & 45.2 & 32.8 &542\\
   &shared int com & 37.9 &35.9 &45.9 & 30.7 &365\\
   &shared int hor  & 26.3&24.7&32.1&  21.7 &227\\
   &additional &1.2 &1.3 &1.2 &0.8 &205\\
   \hline\hline
   $\sqrt{\frac{\log{n_{g_{i}}}\times N}{\log{N}\times n_{g_{i}}}}$&no removal &0 &0 &0&0&0\\
   &all intersection  &23.2 &22.1 & 27.7 & 19.2&131 \\
   &shared int dram &41.8 &43.7 & 43.7 & 33.0 &580\\
   &shared int com & 37.2 &35.6 &45.2 & 29.1 &378\\
   &shared int hor  &26.1 &25.1&31.1&20.9 &232\\
   &additional &1.8 &1.9&1.9 &1.1 &265\\
   \hline\hline
 
 \end{tabular}
 }
 \label{table:6}
\end{table}

\end{document}